\documentclass[sigconf]{acmart}

\usepackage{booktabs} 

\usepackage{amsthm}
\usepackage{amscd,amsfonts,amsbsy,rotating}
\usepackage{makecell}
\usepackage{pifont}
\usepackage{balance}
\usepackage{graphicx}
\usepackage{epsfig,epstopdf}
\usepackage{subfigure}
\usepackage{multirow}
\usepackage{booktabs}
\usepackage{color,xcolor}
\usepackage{url}
\usepackage{latexsym,bm}
\usepackage{enumitem,balance,mathtools}
\usepackage{wrapfig}
\usepackage{euscript}
\usepackage{ifpdf}
\usepackage{diagbox}
\usepackage{caption}
\usepackage{makecell}
\usepackage{subfigure}
\usepackage[leftcaption]{sidecap}
\usepackage{textcomp}
\usepackage[normalem]{ulem}
\usepackage{breakurl}
\usepackage[table]{xcolor}
\usepackage{diagbox}
\usepackage{tcolorbox}

\usepackage{array}
\newcolumntype{N}{@{}m{0pt}@{}}

\usepackage{lipsum}

\PassOptionsToPackage{hyphens}{url}
\usepackage{hyperref}

\usepackage[ruled,linesnumbered]{algorithm2e}

\usepackage{algorithmic}

\usepackage{lipsum}

\newcommand{\minisection}[1]{\vspace{5pt}\noindent\textbf{#1}}

\begin{document}
\title{Latent-Aligned Reasoning for Multimodal Recommendation}

\author{Jiarui Jin}
\affiliation{\institution{Xiaohongshu Inc.}}
\email{jinjiarui@xiaohongshu.com}

\author{Anyang Ji}
\authornote{Work done during internship at Xiaohongshu Inc.}
\affiliation{\institution{Nanjing University}}
\email{jiay@lamda.nju.edu.cn}
\renewcommand{\shortauthors}{Jiarui Jin et al.}
\renewcommand{\shorttitle}{}
\settopmatter{printacmref=false}

\begin{abstract}
Multimodal Vision-Language Models (VLMs) have demonstrated remarkable capabilities in cross-modal understanding, yet a fundamental challenge persists when applying them to recommendation: as representations propagate through multi-step reasoning, both visual and textual signals progressively attenuate — a phenomenon we term \emph{cross-modal dilution}.
To address this, we propose \textbf{\textsc{LARK}} (\textbf{L}atent-\textbf{A}ligned \textbf{R}easoning framewor\textbf{K}), a two-stage latent reasoning framework with complementary alignment mechanisms within a single VLM.
In the first stage, learnable latent tokens are interleaved with multi-step chain-of-thought (CoT) reasoning and explicitly aligned with a frozen vision encoder, serving as visual checkpoints that preserve perceptual details throughout the reasoning chain.
In the second stage, the latent representations are projected via a bridge MLP and trained with item-to-item contrastive learning; to prevent the reasoning semantics from fading, intermediate features are aligned with the CoT hidden states from the first stage, anchoring the final embeddings to the model's own reasoning output.
Experiments on three public benchmarks and one industrial dataset show that \textsc{LARK} achieves state-of-the-art performance across multiple recommendation architectures, with controlled ablations confirming the distinct contribution of each component.
\end{abstract}

\settopmatter{printacmref=false} 
\maketitle


\section{Introduction}
Multimodal recommendations \citep{wei2019mmgcn,wu2022mm,xun2021we,liu2024multimodal} have emerged as an indispensable component of modern recommender systems, enabling personalized item suggestions by capturing users' preferences over rich visual and textual content.
Existing approaches typically extract multimodal information into dense embeddings \citep{zhang2020multimodal} or sparse IDs via lookup tables \citep{luo2025qarm}, relying on modality-specific encoders that process each signal in isolation.
Many efforts have been devoted to enhancing multimodal representations for recommendation by leveraging the cross-modal understanding capabilities of Vision-Language Models (VLMs) \citep{bai2025qwen25vl,liu2024llava}, which have demonstrated remarkable proficiency in tasks such as image captioning \citep{zhang2024overview,cheng2025caparena} and visual question answering \citep{manmadhan2020visual,huynh2025visual}, suggesting a promising paradigm for jointly modeling visual and textual features within a unified architecture.

In parallel, Chain-of-Thought (CoT) reasoning \citep{wei2022chain} has become a cornerstone technique for enhancing the reasoning capabilities of large language models, enabling them to decompose complex tasks into interpretable intermediate steps.
Recent studies \citep{zheng2025deepeyes,li2025imagine,wang2025video} have extended CoT from LLMs to VLMs.
However, in contrast to LLMs \citep{snell2024scaling} where longer reasoning chains yield consistent gains, VLMs frequently struggle with long-context reasoning due to the progressive attenuation of both visual and textual signals as the generated sequence lengthens --- a phenomenon we term \emph{cross-modal dilution}.
As recently demonstrated by \citet{jeon2026vision} and \citet{zhang2025notellm} in the context of visual representations, attention to visual tokens decays substantially as the reasoning chain grows; we observe the same effect extends to textual tokens and is particularly acute in recommendation settings, where VLMs must ultimately produce dense embeddings for downstream models, leaving no opportunity for the auto-regressive process to refresh the fading multimodal signals (see attention analysis in \S\ref{sec:dilution}).

Recent work on latent reasoning \citep{hao2024training,wang2026monet,li2025latent,he2024multi} offers a compelling alternative: instead of decoding intermediate thoughts into discrete tokens, the model reasons in continuous hidden space, bypassing the lossy argmax bottleneck.
However, existing latent reasoning methods target text-generation tasks and have not been explored for recommendation, where the output is a dense embedding rather than a token sequence, and where \emph{both} visual and textual signals --- not vision alone --- must be preserved.

\begin{figure*}[!t]
\centering
\includegraphics[width=0.9\linewidth]{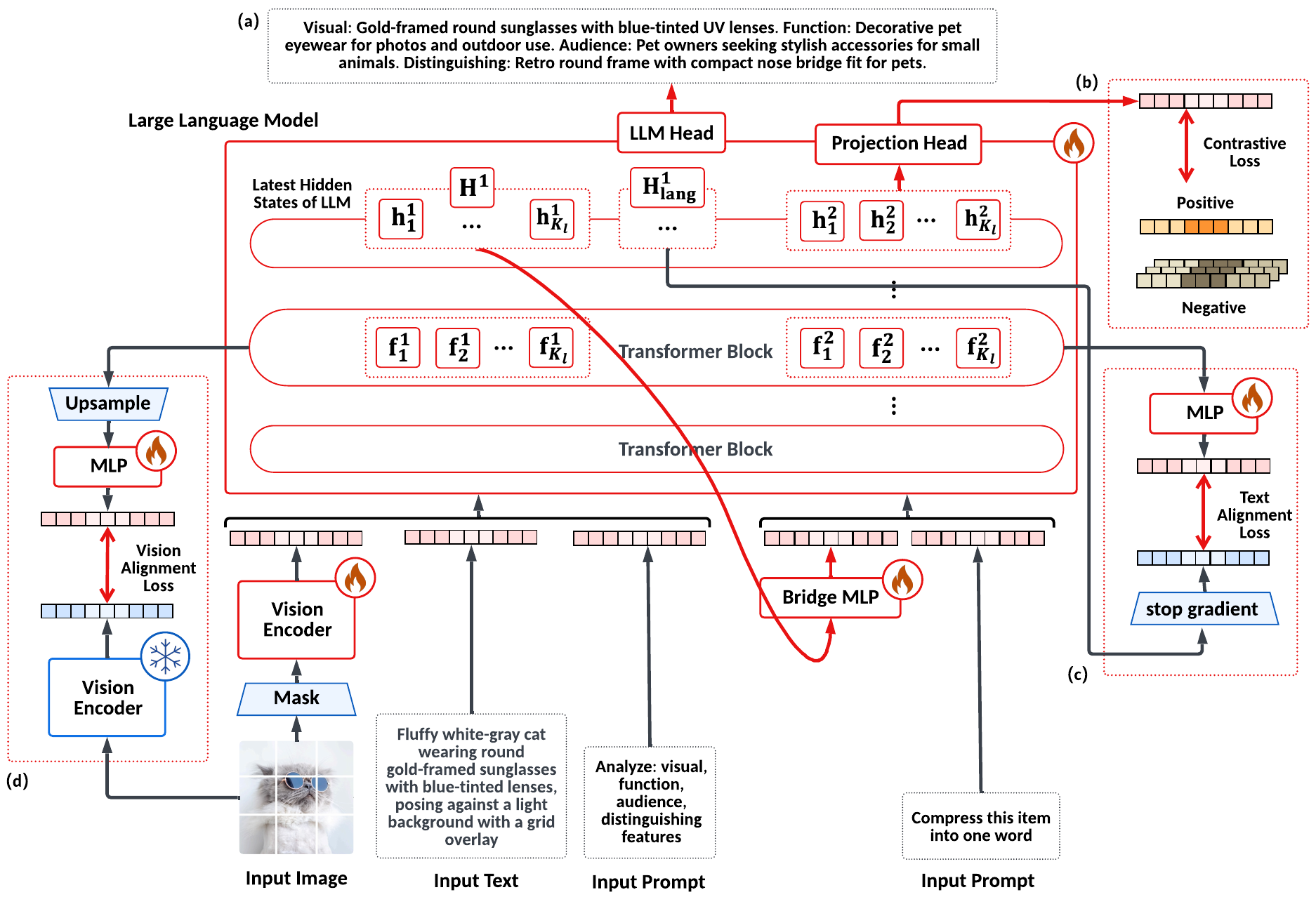}
\caption{
Overview of \textsc{LARK}.
\textbf{Stage~1} interleaves latent tokens with multi-step CoT reasoning; latent groups are aligned patch-wise with a frozen vision encoder~(\textbf{d},~$\mathcal{L}_{\mathrm{vision}}$), while CoT text is supervised by $\mathcal{L}_{\mathrm{cot}}$~(\textbf{a}).
\textbf{Stage~2} projects $\mathbf{H}^{(1)}$ via a Bridge MLP into the same backbone for contrastive training~(\textbf{b},~$\mathcal{L}_{\mathrm{i2i}}$); intermediate features are anchored to the stop-gradient CoT hidden states $\mathbf{H}^{(1)}_{\mathrm{lang}}$~(\textbf{c},~$\mathcal{L}_{\mathrm{text}}$).
}
\label{fig:overview}
\end{figure*}

Building on these advances, we propose \textbf{\textsc{LARK}} (\textbf{L}atent-\textbf{A}ligned \textbf{R}easoning framewor\textbf{K}), which reformulates multimodal recommendation as multi-stage latent reasoning within a single VLM.
\textsc{LARK} operates as a two-stage architecture with complementary alignment mechanisms designed to counteract cross-modal dilution at each stage.
In the first stage, learnable latent tokens are interleaved with multi-step CoT reasoning following a fixed template: before each reasoning step, a group of latent tokens is generated in continuous hidden space and aligned with a frozen vision encoder, serving as visual checkpoints that preserve perceptual information.
In the second stage, the latent representations are projected through a bridge MLP and trained with item-to-item contrastive learning; to prevent the reasoning semantics from fading, intermediate features are aligned with the CoT hidden states from the first stage, anchoring the embeddings to the model's own reasoning output.

We evaluate \textsc{LARK} on three public subsets of the Amazon Reviews 2023 benchmark~\citep{hou2024bridging} and one proprietary dataset from a social media platform serving over 100 million monthly active users.
\textsc{LARK} consistently outperforms all baselines and generalizes across multiple downstream architectures (DeepFM \citep{guo2017deepfm}, LightGCN~\citep{he2020lightgcn}, and SASRec~\citep{kang2018self}).
Controlled ablations isolate the contribution of each component, and attention analyses confirm that cross-modal dilution is effectively mitigated by \textsc{LARK}'s dual alignment mechanisms.

\section{Preliminaries}

\subsection{Problem Formulation}
\label{sec:prelim}
Let $\mathcal{U}$ and $\mathcal{I}$ denote the sets of users and items, respectively.
Each item $i \in \mathcal{I}$ is associated with an image $\mathbf{I}_i$ and a textual description $\mathbf{T}_i = (t_1, t_2, \ldots, t_{|\mathbf{T}_i|})$.
User-item interactions are represented by a binary matrix $\mathbf{R}_{ui} \in \{0, 1\}^{|\mathcal{U}| \times |\mathcal{I}|}$, where $r_{ui} = 1$ indicates that user $u$ has interacted with item $i$.

The goal of multimodal recommendation is to learn a scoring function $\mathcal{F}_\text{score}: \mathcal{U} \times \mathcal{I} \rightarrow \mathbb{R}$ that predicts user preferences over items by jointly leveraging visual and textual content.
Specifically, we aim to learn an item representation function $\mathcal{F}_\text{item}: (\mathbf{I}_i, \mathbf{T}_i) \rightarrow \mathbf{e}_i \in \mathbb{R}^D$ that maps multimodal item content into a dense embedding, and a user representation function $\mathcal{F}_\text{user}: u \rightarrow \mathbf{e}_u \in \mathbb{R}^D$, such that the recommendation score $\mathcal{F}_\text{score}(u, i) = \mathbf{e}_u^\top \mathbf{e}_i$ reflects the likelihood of user $u$ interacting with item $i$.

Following the Bayesian Personalized Ranking (BPR) \citep{rendle2012bpr} paradigm, we optimize the pairwise ranking objective:
\begin{equation}
\label{eqn:bpr}
\mathcal{L}_{\text{BPR}} = -\sum_{(u,i,j) \in \mathcal{D}} \log \sigma\left(\mathbf{e}_u^\top \mathbf{e}_i - \mathbf{e}_u^\top \mathbf{e}_j\right),
\end{equation}
where $(u, i, j)$ denotes a training triple with $r_{ui} = 1$ and $r_{uj} = 0$, and $\sigma(\cdot)$ is the sigmoid function.

Beyond user-item interactions, item-item co-occurrence patterns provide an important source of collaborative signal.
We adopt the Swing algorithm~\citep{yang2020large} to mine item-to-item relevance from user behavior.
For two items $i$ and $j$, let $\mathcal{U}_{ij} = \{u \in \mathcal{U} \mid r_{ui} = 1 \wedge r_{uj} = 1\}$ denote the set of users who have interacted with both.
The Swing similarity is defined as:
\begin{equation}
\label{eqn:swing}
\mathcal{F}_\text{swing}(i, j) = \sum_{u \in \mathcal{U}_{ij}} \sum_{v \in \mathcal{U}_{ij}, v \neq u} \frac{1}{\alpha + |\mathcal{I}_u \cap \mathcal{I}_v|},
\end{equation}
where $\mathcal{I}_u$ denotes the set of items interacted by user $u$, and $\alpha$ is a smoothing constant (set to 1 in implementation).
Intuitively, Swing down-weights co-occurrences contributed by user pairs with highly overlapping histories, which are more likely to reflect popularity bias rather than genuine item-relatedness.
For each item $i$, we retain the top-$K$ items by Swing similarity as its positive item set $\mathcal{N}_i^+$, which serves as the supervision signal for item-to-item contrastive learning in \S~\ref{sec:training}.

\subsection{Related Work}
\minisection{Multi-modal Recommendations.}
Multimodal recommendation leverages visual and textual content to enrich item representations beyond collaborative signals.
Early approaches \citep{wei2019mmgcn,zhang2021mining} integrate pre-extracted multimodal features (e.g., from ResNet \citep{he2016deep} or BERT \citep{devlin2019bert}) into graph-based collaborative filtering frameworks such as LightGCN \citep{he2020lightgcn}.
LATTICE \citep{zhang2021mining} constructs modality-specific item-item graphs to capture latent structures, while BM3 \citep{zhou2023bootstrap} introduces self-supervised multimodal alignment through bootstrap contrastive learning.
More recently, AlignRec \citep{liu2024alignrec} proposes a three-level alignment framework that decomposes the recommendation objective into inter-content, content-category, and user-item alignments.
With the rise of LLMs, several studies \citep{wei2024llmrec,zhang2025notellm} have begun exploring VLMs for item representation learning, demonstrating that unified vision-language architectures can capture richer cross-modal semantics than modality-specific encoders.
Meanwhile, \citet{jin2025not} propose a multiple-round recommender system that jointly optimizes queries and items, highlighting the importance of iterative reasoning in recommendation.
However, these approaches typically treat VLMs as static feature extractors or fine-tune them with standard next-token prediction objectives, without incorporating chain-of-thought reasoning to enable progressive multimodal understanding. 
As a result, they fail to address the cross-modal dilution problem, where both visual and textual signals attenuate through multi-step reasoning.

\minisection{Chain-of-Thought and Latent Reasoning.}
Chain-of-Thought (CoT) prompting \citep{wei2022chain} has proven effective in improving the reasoning capabilities of LLMs by decomposing complex problems into intermediate textual steps.
Subsequent work has extended CoT to multimodal settings \citep{zheng2025deepeyes,li2025imagine,wang2025video}, enabling VLMs to reason over visual inputs.
However, standard CoT reasoning in VLMs suffers from the progressive attenuation of visual signals as the reasoning chain lengthens \citep{snell2024scaling}, since the auto-regressive process only refreshes textual tokens while visual information fades.
To address this, latent reasoning approaches such as Coconut \citep{hao2024training} propose reasoning entirely in continuous hidden space, bypassing the lossy argmax decoding at each step.
VaLR \citep{jeon2026vision} extends this idea to multimodal models by interleaving vision-aligned latent tokens with CoT reasoning steps, where each group of latent tokens is explicitly grounded in external vision encoder features via representation alignment (REPA) \citep{yu2024representation}..
In contrast, multimodal recommendation requires preserving \emph{both} visual and textual signals, since the output is a dense embedding where neither modality is self-refreshing.
\textsc{LARK} bridges this gap by introducing complementary alignment mechanisms: vision alignment in the first reasoning stage to preserve perceptual details, and reasoning-text alignment in the second stage to retain semantic understanding from the CoT process.

\section{\textsc{LARK}}
\subsection{Architecture of \textsc{LARK}}
We propose \textsc{LARK}, a framework that enables multimodal recommendation to conduct latent reasoning before producing the final item representation.
As illustrated in Figure~\ref{fig:overview}, \textsc{LARK} operates as a two-stage architecture with complementary alignment mechanisms within a single VLM, connected by a Bridge MLP that projects latent representations across reasoning stages.

The autoregressive generation process operates in two modes: \emph{latent mode} and \emph{language mode}.
In latent mode, the model's hidden state is directly fed back as the input embedding for the next position, enabling reasoning in continuous space without discretization.
In language mode, the model generates text tokens through the LLM head, producing interpretable chain-of-thought reasoning steps.
To delineate these two modes, we introduce special \texttt{<latent>} and \texttt{</latent>} tokens into the vocabulary.
During training, latent and language segments are arranged according to a fixed template: before each CoT reasoning step, $K$ latent tokens are inserted, yielding an interleaved sequence of the form
\begin{equation}
\label{eqn:interleave}
\underbrace{\ell_1, \ldots, \ell_K}_{\text{latent}},\; \underbrace{r^{(1)}_1, \ldots, r^{(1)}_{n_1}}_{\text{CoT step 1}},\; \underbrace{\ell_1, \ldots, \ell_K}_{\text{latent}},\; \underbrace{r^{(2)}_1, \ldots, r^{(2)}_{n_2}}_{\text{CoT step 2}},\; \ldots
\end{equation}
where each latent segment uses hidden-state feedback (latent mode) and each CoT segment uses standard token embeddings (language mode), supervised by teacher-generated multi-step reasoning annotations (\S\ref{sec:training}).

This design enables two complementary alignment mechanisms to counteract cross-modal dilution at each stage:
(i) \textbf{Vision-Language Reasoning Stage.} Each group of latent tokens in the first stage is aligned with a frozen vision encoder, preserving perceptual details throughout the reasoning chain.
(ii) \textbf{Item-Item Alignment Stage.} The second stage's intermediate features are aligned with the CoT hidden states from Stage~1, anchoring the item embeddings to the model's own reasoning output.

\textsc{LARK} serves as an offline multimodal item encoder: item embeddings are precomputed and stored in a vector index before deployment.
The downstream recommendation model (e.g., LightGCN~\citep{he2020lightgcn}, SASRec~\citep{kang2018self}) consumes these frozen item embeddings to learn user representations and is updated independently during online serving.
This decoupled design allows \textsc{LARK} to scale to large item catalogs without introducing latency at serving time.

\subsection{Vision-Language Reasoning Stage}
\label{sec:vl_align}

The first stage fuses visual and textual information into a unified latent representation through interleaved latent-language reasoning.

Given the $i$-th item with input image $\mathbf{I}_i$ and textual description $\mathbf{T}_i$, we construct token-level embeddings.
A masking strategy randomly masks a fraction $\rho$ of the image patches before encoding (where $\rho$ is set to 50\% in implementation); the trainable vision encoder then processes the remaining visible patches to yield vision tokens $\mathbf{V}_i = (\mathbf{v}_1, \mathbf{v}_2, \ldots, \mathbf{v}_V)$.
The text is tokenized and embedded as $\mathbf{W}_i = (\mathbf{w}_1, \mathbf{w}_2, \ldots, \mathbf{w}_T)$.
A task-specific prompt is appended with token embeddings $\mathbf{P}^{(1)}=(\mathbf{p}^{(1)}_1, \mathbf{p}^{(1)}_2,\ldots, \mathbf{p}^{(1)}_P)$.
The complete input sequence $\mathbf{E}^{(1)}$ for the first stage is:
\begin{equation}
\label{eqn:input_stage1}
\mathbf{E}^{(1)} = [\mathbf{v}_1, \ldots, \mathbf{v}_V;\; \mathbf{w}_1, \ldots, \mathbf{w}_T;\; \mathbf{p}^{(1)}_1, \ldots, \mathbf{p}^{(1)}_P],
\end{equation}
where $[\,\cdot\,;\,\cdot\,]$ denotes concatenation.

After processing $\mathbf{E}^{(1)}$, the transformer conducts autoregressive reasoning by interleaving $N$ groups of $K$ latent steps with $N$ multi-step CoT reasoning segments, as described in Eq.~(\ref{eqn:interleave}).
The total number of latent tokens is $K_l = N \times K$.
Let $\mathbf{S}_0 = \mathbf{E}^{(1)}$ denote the initial context.
At each step $t$, the transformer produces a hidden state $\mathbf{h}_t$:
\begin{equation}
\mathbf{h}_t = \mathcal{F}_\text{transformer}(\mathbf{S}_t)[-1,:],
\end{equation}
and the context is extended as $\mathbf{S}_{t+1} = [\mathbf{S}_t;\; \mathbf{e}_{t+1}]$, where $\mathbf{e}_{t+1}$ depends on whether position $t+1$ falls in a latent or language segment according to the fixed template:
\begin{equation}
\label{eqn:dual_mode}
\mathbf{e}_{t+1} =
\begin{cases}
    \mathbf{h}_t & \text{if } t+1 \in \mathcal{T}_{\text{latent}} \quad (\text{latent mode}), \\
    \mathcal{F}_\text{embed}(\mathbf{c}^*_{t+1}) & \text{if } t+1 \in \mathcal{T}_{\text{lang}} \quad (\text{language mode}),
\end{cases}
\end{equation}
where $\mathcal{T}_{\text{latent}}$ and $\mathcal{T}_{\text{lang}}$ denote the sets of latent and language positions defined by the template, and $\mathbf{c}^*_{t+1}$ is the ground-truth CoT token at position $t+1$ (teacher forcing during training).
In latent mode, the hidden state is directly recycled as the next input, preserving the continuous representation without discretization.
In language mode, the generated tokens form multi-step CoT reasoning about the item (e.g., visual attributes, functional analysis, target audience, and distinguishing features), supervised by cross-entropy loss during training (Eq.~(\ref{eqn:cot_loss}) in \S\ref{sec:training}).
Thanks to causal attention, each latent token can attend to the entire preceding context --- including input tokens and previously generated CoT tokens --- enabling effective compression of multimodal information.

\minisection{Vision Alignment.}
To counteract cross-modal dilution across reasoning stages, we
apply representation alignment to the first
stage's intermediate features.
Let $\mathbf{F}^{(1)} = (\mathbf{f}_1^{(1)}, \mathbf{f}_2^{(1)}, \ldots,
\mathbf{f}_{K_l}^{(1)})$ denote the intermediate representations at
layer $m = \lfloor 2L/3 \rfloor$ of the transformer, where $L$
is the total number of layers.
A frozen copy of the VLM's own vision encoder processes the unmasked input image to produce patch-wise features $\mathbf{F}_\phi = \mathcal{F}_\text{frozen}(\mathbf{I}_i)\in \mathbb{R}^{N_p \times D_\phi}$, serving as the alignment target.
The intermediate features are upsampled and projected through a learnable MLP $\mathcal{F}_\text{MLP}^{\text{vis}}$:
$\hat{\mathbf{F}}^{(1)} = \mathcal{F}_\text{MLP}^{\text{vis}}(\text{Upsample}(\mathbf{F}^{(1)}))$.

The alignment loss is designed to maximize the patch-wise cosine similarity between the projected features and the frozen encoder's output:
\begin{equation}
\label{eqn:vis_align}
\mathcal{L}_{\text{vision}} = - \frac{1}{N_p} \sum_{j=1}^{N_p}
\cos\!\left(\hat{\mathbf{f}}_j^{(1)},\; \mathbf{f}_{\phi,j}\right),
\end{equation}
where $\cos(\cdot, \cdot)$ denotes cosine similarity, and $\mathbf{f}_{\phi,j} = \mathbf{F}_\phi[j, :]$.
This alignment anchors the first stage's latent representations to a semantically rich
visual space, preventing them from drifting
as they propagate through the reasoning chain.

\minisection{Hidden State Collection.}
The first stage collects the last-layer hidden states at all $K_l$ latent
token positions, re-indexed as $k = 1, \ldots, K_l$:
$\mathbf{H}^{(1)} = (\mathbf{h}_1, \mathbf{h}_2, \ldots, \mathbf{h}_{K_l})$,
which encodes a vision-language fused representation of the item.
It also collects the hidden states at language positions as the CoT representation:
$\mathbf{H}^{(1)}_\text{lang} = (\mathbf{h}_{t})_{t \in \mathcal{T}_{\text{lang}}}$,
which is used as the alignment target in the second stage (\S\ref{sec:ii_align}).
Both $\mathbf{H}^{(1)}$ and $\mathbf{H}^{(1)}_\text{lang}$ are collected from the first stage; $\mathbf{H}^{(1)}$ is forwarded via the Bridge MLP to the second stage, while $\mathbf{H}^{(1)}_\text{lang}$ serves as the alignment target (with stop-gradient) in \S\ref{sec:ii_align}.

\subsection{Item-Item Alignment Stage}
\label{sec:ii_align}
The second stage derives compact item embeddings suitable for recommendation by grounding the latent representations in item-level relational structure.

The hidden states $\mathbf{H}^{(1)}$ from the first stage reside in the transformer's hidden-state space, which differs distributionally from the embedding space expected by the transformer's first layer.
To bridge this gap, we project $\mathbf{H}^{(1)}$ through a learnable Bridge MLP:
\begin{equation}
\label{eqn:bridge}
\tilde{\mathbf{H}}^{(1)} = \mathcal{F}_\text{BridgeMLP}(\mathbf{H}^{(1)}),
\end{equation}
where $\mathcal{F}_\text{BridgeMLP}(\cdot)$ maps from the hidden-state space to the embedding space, enabling seamless integration with the second-stage input.
The second-stage input is constructed by concatenating the bridged representations with a task prompt encoded as $\mathbf{P}^{(2)}=(\mathbf{p}^{(2)}_1, \mathbf{p}^{(2)}_2,\ldots, \mathbf{p}^{(2)}_Q)$:
\begin{equation}
\label{eqn:input_stage2}
\mathbf{E}^{(2)} = [\tilde{\mathbf{h}}_1, \ldots, \tilde{\mathbf{h}}_{K_l};\; \mathbf{p}^{(2)}_1, \ldots, \mathbf{p}^{(2)}_Q].
\end{equation}
Since $\tilde{\mathbf{h}}$s consist of continuous vectors rather than discrete tokens, the second stage operates entirely in the embedding space, bypassing tokenization.

\minisection{Reasoning-Text Alignment.}
To prevent the reasoning semantics captured in the first stage from fading during the second stage, we align the second stage's intermediate features with the CoT hidden states produced by the first stage.
Let $\mathbf{F}^{(2)} = (\mathbf{f}_1^{(2)}, \mathbf{f}_2^{(2)}, \ldots,
\mathbf{f}_{K_l}^{(2)})$ denote the intermediate representations at
layer $m = \lfloor 2L/3 \rfloor$ of the transformer, where $L$
is the total number of layers.
The alignment target is the mean-pooled CoT representation from Stage~1:
$\bar{\mathbf{r}} = \frac{1}{|\mathcal{T}_{\text{lang}}|}\sum_{t \in \mathcal{T}_{\text{lang}}} \mathbf{h}_t$,
computed from the language-position hidden states $\mathbf{H}^{(1)}_\text{lang}$ collected in \S\ref{sec:vl_align}.
A stop-gradient operator is applied to $\bar{\mathbf{r}}$ to prevent the alignment loss from back-propagating into Stage~1's language generation.
The reasoning-text alignment loss is:
\begin{equation}
\label{eqn:text_align}
\mathcal{L}_{\text{text}} = - \frac{1}{K_l} \sum_{j=1}^{K_l}
\cos\!\left(\mathcal{F}_\text{MLP}^{\text{text}}(\mathbf{f}_j^{(2)}),\; \text{sg}(\bar{\mathbf{r}})\right),
\end{equation}
where $\mathcal{F}_\text{MLP}^{\text{text}}$ is a learnable projection and $\text{sg}(\cdot)$ denotes stop-gradient.
This alignment anchors the second stage to the semantic understanding produced by the first stage's reasoning process, ensuring that the reasoning conclusions are preserved in the final item embeddings.
Note that while all latent positions are aligned to the same pooled target $\bar{\mathbf{r}}$, token-level diversity is maintained by two mechanisms: (i) the contrastive loss $\mathcal{L}_\text{i2i}$ (as later formulated in Eq.~(\ref{eqn:contrastive})) requires the final embeddings to discriminate between items, which propagates discriminative gradients back through $\mathbf{F}^{(2)}$; (ii) causal attention ensures that each latent position attends to a different prefix of the input, producing naturally heterogeneous intermediate features.

The transformer processes $\mathbf{E}^{(2)}$ to produce the final hidden states $\mathbf{H}^{(2)}$.
The item embedding is obtained through a projection head, denoted as $\mathcal{F}_\text{ProjectionHead}$:
\begin{equation}
\label{eqn:item_emb}
\mathbf{e}_i = \mathcal{F}_\text{ProjectionHead}(\mathbf{H}^{(2)}),
\end{equation}
where $\mathcal{F}_\text{ProjectionHead}(\cdot)$ is a learnable MLP that mean-pools the hidden states and maps them to a $D$-dimensional embedding $\mathbf{e}_i \in \mathbb{R}^D$.

The item embeddings are trained with an InfoNCE contrastive loss~\citep{oord2018representation} using item-to-item collaborative signals derived from Swing similarity (\S\ref{sec:prelim}).
For each item $i$, we sample a positive item $j \in \mathcal{N}_i^+$ (top-$K$ Swing neighbors, as derived in \S\ref{sec:prelim}) and treat other in-batch items as negatives:
\begin{equation}
\label{eqn:contrastive}
\mathcal{L}_{\text{i2i}} = -\log \frac{\exp(\cos(\mathbf{e}_i, \mathbf{e}_j) / \tau)}{\sum_{k=1}^{B} \exp(\cos(\mathbf{e}_i, \mathbf{e}_k) / \tau)},
\end{equation}
where $\tau$ is the temperature hyperparameter and $B$ is the batch size.

\begin{figure*}[t!]
\centering
\includegraphics[width=0.8\linewidth]{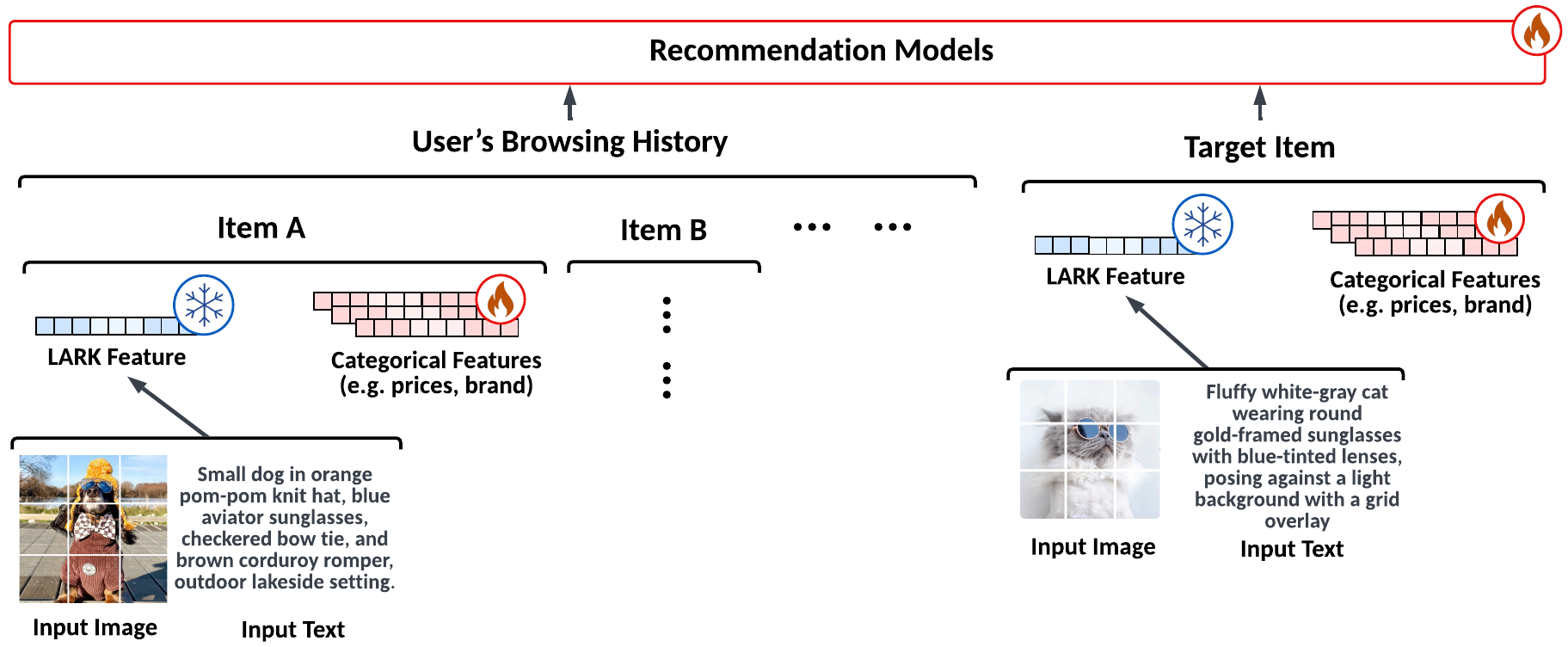}
\caption{
Deployment of \textsc{LARK} in a recommendation pipeline.
\textsc{LARK} encodes each item's image and text into a dense embedding offline (frozen; indicated by the snowflake icon).
The precomputed \textsc{LARK} features are concatenated with categorical features (\emph{e.g.}, price, brand; trainable, indicated by the flame icon) and fed into downstream recommendation models such as DeepFM or LightGCN, which are trained independently on user-item interactions.
}
\label{fig:recommendation}
\end{figure*}

\subsection{Training and Inference}
\label{sec:training}

\textsc{LARK} is trained end-to-end with four complementary objectives:

\minisection{CoT Generation Loss.}
In the first stage, the language-mode outputs are supervised by cross-entropy loss against multi-step reasoning annotations $\mathbf{c}_i^*$, which are generated offline by a teacher VLM (Gemini-2.5-Flash~\citep{gemini2025flash} in our experiments):
\begin{equation}
\label{eqn:cot_loss}
\mathcal{L}_{\text{cot}} = -\sum_{t \in \mathcal{T}_{\text{lang}}}
\log P(\mathbf{c}^*_t \mid \mathbf{E}_{<t}),
\end{equation}
where $\mathcal{T}_{\text{lang}}$ denotes the set of language-mode positions.

\minisection{Vision Alignment Loss.}
The vision alignment loss $\mathcal{L}_{\text{vision}}$ (Eq.~(\ref{eqn:vis_align})) aligns the first stage's latent representations with the frozen vision encoder, preserving visual fidelity across reasoning steps.

\minisection{Reasoning-Text Alignment Loss.}
The reasoning-text alignment loss $\mathcal{L}_{\text{text}}$ (Eq.~(\ref{eqn:text_align})) aligns the second stage's intermediate features with the stop-gradient CoT hidden states from the first stage, anchoring the item embeddings to the reasoning semantics.

\minisection{Contrastive Loss.}
The contrastive loss $\mathcal{L}_{\text{i2i}}$ (Eq.~(\ref{eqn:contrastive})) trains item embeddings using Swing-derived collaborative signals.

The overall training objective is:
\begin{equation}
\label{eqn:total_loss}
\mathcal{L} =\mathcal{L}_{\text{i2i}} + \lambda_\text{cot} \mathcal{L}_{\text{cot}} + \lambda_\text{align} (\mathcal{L}_{\text{vision}} + \mathcal{L}_{\text{text}}),
\end{equation}
where $\lambda_\text{cot}$ and $\lambda_\text{align}$ are balancing hyperparameters.

\minisection{Plug-into Recommendation Models.}
As illustrated in Figure~\ref{fig:recommendation}, \textsc{LARK} serves as a general-purpose item encoder that plugs into arbitrary downstream recommendation models.
For each item $i$, the precomputed embedding $\mathbf{e}_i$ is concatenated with categorical features (e.g., price, brand) to form the item representation.
Given a user $u$ with browsing history $(i_1, i_2, \ldots, i_n)$, the downstream model $\mathcal{F}_{\text{RecommendationModel}}$ aggregates historical item representations into a user embedding $\mathbf{e}_u = \mathcal{F}_{\text{RecommendationModel}}(\mathbf{e}_{i_1}, \mathbf{e}_{i_2}, \ldots, \mathbf{e}_{i_n})$ and scores candidate items via inner product $\hat{y}_{ui} = \mathbf{e}_u^\top \mathbf{e}_i$.
The recommendation model is trained with the BPR loss~\citep{rendle2009bpr}:
\begin{equation}
\label{eqn:bpr_loss}
\mathcal{L}_{\text{rec}} = -\sum_{(u,i,j) \in \mathcal{O}} \log \sigma(\hat{y}_{ui} - \hat{y}_{uj}),
\end{equation}
where $\mathcal{O} = \{(u,i,j) \mid i \in \mathcal{I}_u^+,\, j \notin \mathcal{I}_u^+\}$ denotes the set of pairwise training triplets, $\mathcal{I}_u^+$ is the set of items user $u$ has interacted with, and $\sigma(\cdot)$ is the sigmoid function.
Since \textsc{LARK} embeddings are frozen during this stage, the design is agnostic to the choice of $\mathcal{F}_{\text{RecommendationModel}}$ -- we validate with table-based (DeepFM~\citep{guo2017deepfm}), graph-based (LightGCN~\citep{he2020lightgcn}), and sequential (SASRec~\citep{kang2018self}) architectures in our experiments.

\minisection{Inference and Deployment.}
At inference, the first stage runs the complete latent-language interleaving, generating both latent tokens and CoT text autoregressively.
The resulting $\mathbf{H}^{(1)}$ is then passed through the Bridge MLP to the second stage, which produces the final item embedding $\mathbf{e}_i$.

As depicted in Figure~\ref{fig:recommendation}, all item embeddings can be precomputed offline and stored in a vector index.
At serving time, the downstream recommendation model (e.g., DeepFM, LightGCN or SASRec) consumes these frozen item embeddings to learn and update user representations.
This decoupled design ensures that \textsc{LARK}'s two-stage reasoning incurs no additional latency during online serving, while the downstream model can be updated independently as user behavior evolves.

\begin{figure}[t]
\centering
\includegraphics[width=\linewidth]{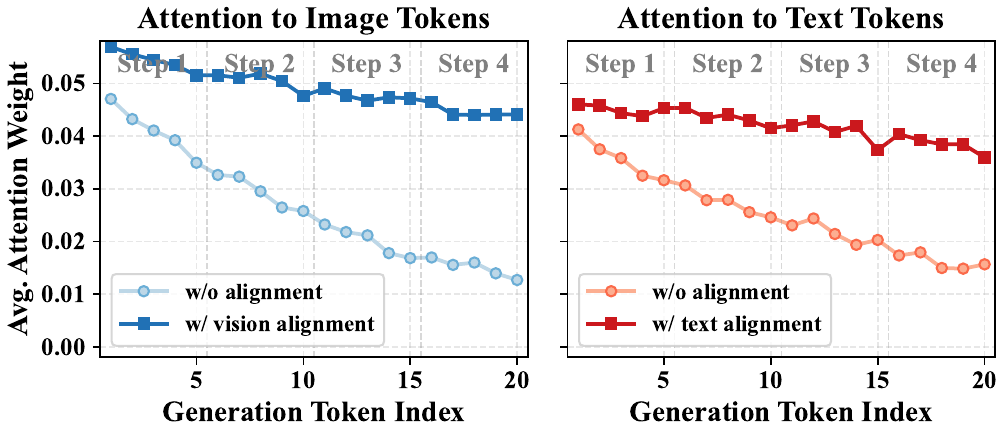}
\vspace{-6mm}
\caption{Attention weight to image tokens (left) and text tokens (right) across reasoning steps in Stage~1 on the Baby dataset. Without alignment, attention to both modalities decays as reasoning progresses; \textsc{LARK}'s dual alignment substantially mitigates this cross-modal dilution.}
\label{fig:attention_decay}
\end{figure}

\section{Experiments}
\subsection{Experimental Setup}
\minisection{Dataset Description.}
We evaluate \textsc{LARK} on three public subsets of the Amazon Reviews benchmark~\citep{he2016ups} -- \textbf{Baby}, \textbf{Sports}, and \textbf{Clothing} -- and one proprietary dataset (\textbf{In-House}) collected from a social media platform serving over 100 million monthly active users.
For Amazon datasets, we apply 5-core filtering following prior work~\citep{zhang2021mining,zhou2023bootstrap}.
Table~\ref{tab:datasets} summarizes the dataset statistics.
For all datasets, we follow the temporal leave-one-out protocol: for each user, the last interacted item is held out for testing, the second-to-last for validation, and the remaining for training.
The Swing item-item co-occurrence graph used for contrastive supervision (\S\ref{sec:prelim}) is constructed exclusively from training-set interactions, ensuring no test-set leakage into the positive neighbor sets $\mathcal{N}_i^+$.

\minisection{Baseline Description.}
We compare \textsc{LARK} with representative methods spanning three categories:
(i) general collaborative filtering (BPR~\citep{rendle2009bpr}, DeepFM~\citep{guo2017deepfm}, LightGCN~\citep{he2020lightgcn}, SASRec~\citep{kang2018self});
(ii) LLM-augmented (LLMRec~\citep{wei2024llmrec}, RLMRec~\citep{ren2024rlmrec});
(iii) multi-modal (VBPR~\citep{he2016vbpr}, MMGCN~\citep{wei2019mmgcn}, LATTICE~\citep{zhang2021mining}, BM3~\citep{zhou2023bootstrap}, FREEDOM~\citep{zhou2023freedom}, AlignRec~\citep{liu2024alignrec}, NoteLLM-2~\citep{zhang2025notellm}).
NoteLLM-2 is originally designed for item-to-item retrieval; for fair comparison, we feed its embeddings into LightGCN following the same protocol as \textsc{LARK}.

As described in \S\ref{sec:training}, we further evaluate \textsc{LARK} by plugging its frozen item embeddings into three representative downstream models: \textsc{LARK}\textsubscript{DeepFM}, \textsc{LARK}\textsubscript{LightGCN}, and \textsc{LARK}\textsubscript{SASRec}, to verify that the learned representations generalize across different recommendation architectures.

\minisection{Evaluation Metric.}
We adopt the all-ranking protocol and report Recall@$K$ (R@$K$) and NDCG@$K$ (N@$K$) with $K \in \{10, 20\}$.
All results are averaged over five runs with different random seeds.

\minisection{Implementation Details.}
\textsc{LARK} is instantiated with Qwen2.5-VL-3B~\citep{bai2025qwen25vl} as the VLM backbone.
We optimize with AdamW using a learning rate of $1\times10^{-4}$ and cosine annealing.
The loss weights in Eq.~(\ref{eqn:total_loss}) are set to $\lambda_\text{cot} = \lambda_\text{align} = 0.1$.
The number of latent tokens per group is $K = 4$ with $N = 4$ reasoning steps, yielding a total of $K_l = N \times K = 16$ latent tokens, and the masking ratio is $\rho = 0.5$.
For Swing-based neighbor mining, we set the smoothing constant $\alpha = 1$ with top-$K = 10$ neighbors, and the contrastive temperature $\tau = 0.07$.
We use Gemini-2.5-Flash~\citep{gemini2025flash} as the teacher VLM to generate multi-step reasoning annotations for all items offline as the CoT supervision.
All codes and datasets would be public upon publication.

\begin{table}[t]
\centering
\caption{Dataset statistics. Amazon datasets are 5-core filtered.}
\label{tab:datasets}
\vspace{-3mm}
\resizebox{0.8\columnwidth}{!}{
\begin{tabular}{lrrrr}
\toprule
\textbf{Dataset} & \textbf{\#Users} & \textbf{\#Items} & \textbf{\#Inter.} & \textbf{Density} \\
\midrule
Baby     & 19,445  & 7,050   & 139,110    & 0.101\% \\
Sports   & 35,598  & 18,357  & 296,337    & 0.045\% \\
Clothing & 39,387  & 23,033  & 278,677    & 0.031\% \\
In-House & 856,214 & 432,671 & 12,384,529 & 0.003\% \\
\bottomrule
\vspace{-3mm}
\end{tabular}
}
\end{table}

\begin{table*}[!t]
\centering
\caption{Overall performance comparison. Best in \textbf{bold}, second-best \underline{underlined}. $*$ indicates statistically significant improvement over the best baseline (paired $t$-test, $p < 0.05$).
Some baseline results are taken from \citet{zhou2023freedom} and \citet{liu2024alignrec} where available, and reproduced using official implementations otherwise.}
\vspace{-3mm}
\label{tab:main}
\resizebox{\textwidth}{!}{
\begin{tabular}{lcccccccccccccccc}
\toprule
\multirow{2}[2]{*}{\textbf{Method}}
  & \multicolumn{4}{c}{\textbf{Baby}}
  & \multicolumn{4}{c}{\textbf{Sports}}
  & \multicolumn{4}{c}{\textbf{Clothing}}
  & \multicolumn{4}{c}{\textbf{In-House}} \\
\cmidrule(lr){2-5}\cmidrule(lr){6-9}\cmidrule(lr){10-13}\cmidrule(lr){14-17}
 & R@10 & R@20 & N@10 & N@20
 & R@10 & R@20 & N@10 & N@20
 & R@10 & R@20 & N@10 & N@20
 & R@10 & R@20 & N@10 & N@20 \\
\midrule
\rowcolor[HTML]{f0f0f0}
\multicolumn{17}{c}{\textit{\textbf{General Collaborative Filtering}}} \\
BPR
 & .0357 & .0575 & .0192 & .0249
 & .0432 & .0653 & .0241 & .0298
 & .0206 & .0303 & .0114 & .0138
 & .0187 & .0312 & .0098 & .0131 \\
DeepFM
 & .0438 & .0685 & .0235 & .0301
 & .0512 & .0781 & .0281 & .0351
 & .0318 & .0482 & .0172 & .0214
 & .0205 & .0342 & .0109 & .0145 \\
LightGCN
 & .0479 & .0754 & .0257 & .0328
 & .0569 & .0864 & .0311 & .0387
 & .0361 & .0544 & .0197 & .0243
 & .0221 & .0365 & .0118 & .0155 \\
SASRec
 & .0465 & .0732 & .0248 & .0318
 & .0582 & .0885 & .0318 & .0396
 & .0348 & .0528 & .0189 & .0235
 & .0228 & .0378 & .0122 & .0160 \\
\midrule
\rowcolor[HTML]{f0f0f0}
\multicolumn{17}{c}{\textit{\textbf{LLM-based Recommendation}}} \\
LLMRec
 & .0538 & .0841 & .0288 & .0367
 & .0625 & .0948 & .0339 & .0422
 & .0498 & .0741 & .0271 & .0334
 & .0255 & .0413 & .0137 & .0178 \\
RLMRec
 & .0552 & .0862 & .0295 & .0377
 & .0638 & .0968 & .0346 & .0431
 & .0508 & .0758 & .0276 & .0342
 & .0260 & .0421 & .0140 & .0182 \\
\midrule
\rowcolor[HTML]{f0f0f0}
\multicolumn{17}{c}{\textit{\textbf{Multi-Modal Recommendation}}} \\
VBPR
 & .0385 & .0594 & .0201 & .0258
 & .0438 & .0668 & .0237 & .0298
 & .0218 & .0336 & .0117 & .0148
 & .0198 & .0335 & .0105 & .0141 \\
MMGCN
 & .0421 & .0660 & .0220 & .0282
 & .0401 & .0636 & .0209 & .0270
 & .0227 & .0361 & .0120 & .0154
 & .0235 & .0382 & .0126 & .0164 \\
LATTICE
 & .0547 & .0850 & .0292 & .0370
 & .0620 & .0953 & .0335 & .0421
 & .0492 & .0733 & .0268 & .0330
 & .0258 & .0415 & .0139 & .0180 \\
BM3
 & .0564 & .0883 & .0301 & .0383
 & .0656 & .0980 & .0355 & .0438
 & .0577 & .0845 & .0316 & .0387
 & .0271 & .0438 & .0147 & .0190 \\
FREEDOM
 & .0627 & .0992 & .0330 & .0424
 & .0717 & .1089 & .0385 & .0481
 & .0629 & .0941 & .0341 & .0420
 & .0288 & .0462 & .0158 & .0203 \\
AlignRec
 & \underline{.0674} & \underline{.1046} & \underline{.0363} & \underline{.0458}
 & .0758 & .1160 & .0414 & .0517
 & \underline{.0668} & \underline{.1002} & \underline{.0362} & \underline{.0447}
 & \underline{.0312} & \underline{.0498} & \underline{.0171} & \underline{.0219} \\
NoteLLM-2
 & .0661 & .1028 & .0355 & .0448
 & \underline{.0771} & \underline{.1175} & \underline{.0421} & \underline{.0526}
 & .0652 & .0972 & .0354 & .0435
 & .0298 & .0478 & .0163 & .0210 \\
\midrule
\rowcolor[HTML]{ecf0ff}
\textbf{\textsc{LARK}\textsubscript{DeepFM}}
 & .0712 & .1098 & .0383 & .0479
 & .0813 & .1232 & .0441 & .0548
 & .0695 & .1031 & .0374 & .0459
 & .0328 & .0519 & .0179 & .0230 \\
\rowcolor[HTML]{ecf0ff}
\textbf{\textsc{LARK}\textsubscript{LightGCN}}
 & \textbf{.0741}$^*$ & \textbf{.1148}$^*$ & \textbf{.0402}$^*$ & \textbf{.0506}$^*$
 & \textbf{.0842}$^*$ & \textbf{.1285}$^*$ & \textbf{.0462}$^*$ & \textbf{.0576}$^*$
 & \textbf{.0738}$^*$ & \textbf{.1108}$^*$ & \textbf{.0403}$^*$ & \textbf{.0498}$^*$
 & \textbf{.0352}$^*$ & \textbf{.0548}$^*$ & \textbf{.0194}$^*$ & \textbf{.0248}$^*$ \\
\rowcolor[HTML]{ecf0ff}
\textbf{\textsc{LARK}\textsubscript{SASRec}}
 & .0729 & .1131 & .0396 & .0498
 & .0825 & .1256 & .0451 & .0563
 & .0721 & .1079 & .0393 & .0484
 & .0341 & .0535 & .0187 & .0240 \\
\bottomrule
\end{tabular}
}
\end{table*}

\subsection{Performance Comparisons}
Table~\ref{tab:main} presents the overall performance comparison.
We summarize our findings as follows:

\textsc{LARK} consistently outperforms all baselines across four datasets and all metrics, regardless of the downstream recommendation model.
On the large-scale In-House dataset, \textsc{LARK}\textsubscript{LightGCN} achieves R@20 of 0.0548, improving over AlignRec by 10.0\%, demonstrating its scalability to industrial settings.
Note that \textsc{LARK} (3B) uses a smaller backbone than NoteLLM-2~\citep{zhang2025notellm} (7B), yet consistently outperforms it, demonstrating that the two-stage latent reasoning design is more parameter-efficient than naive LLM-based approaches.

Multi-modal methods substantially outperform pure collaborative filtering methods, confirming the value of incorporating visual and textual signals and the effectiveness of cross-modal alignment in capturing item semantics beyond interaction patterns alone.
Among them, AlignRec achieves the best baseline performance on most datasets through explicit multi-level alignment, while NoteLLM-2 shows competitive or superior results on Sports by leveraging LLM-derived multimodal representations.
Also, LLMRec, which augments collaborative filtering with LLM-generated text without directly utilizing visual features, performs below dedicated multi-modal methods, indicating that textual augmentation alone cannot substitute for explicit visual modeling.

\begin{table*}[!t]
\centering
\caption{Ablation study of \textsc{LARK}\textsubscript{LightGCN} across all datasets.}
\vspace{-3mm}
\label{tab:ablation}
\resizebox{\textwidth}{!}{
\begin{tabular}{lcccccccccccccccc}
\toprule
\multirow{2}[2]{*}{\textbf{Variant}}
  & \multicolumn{4}{c}{\textbf{Baby}}
  & \multicolumn{4}{c}{\textbf{Sports}}
  & \multicolumn{4}{c}{\textbf{Clothing}}
  & \multicolumn{4}{c}{\textbf{In-House}} \\
\cmidrule(lr){2-5}\cmidrule(lr){6-9}\cmidrule(lr){10-13}\cmidrule(lr){14-17}
 & R@10 & R@20 & N@10 & N@20
 & R@10 & R@20 & N@10 & N@20
 & R@10 & R@20 & N@10 & N@20
 & R@10 & R@20 & N@10 & N@20 \\
\midrule
\rowcolor[HTML]{f0f0f0}
\multicolumn{17}{c}{\textit{\textbf{Effect of CoT Supervision}}} \\
w/o CoT
 & .0698 & .1078 & .0377 & .0475
 & .0795 & .1213 & .0435 & .0543
 & .0697 & .1045 & .0380 & .0469
 & .0331 & .0516 & .0182 & .0233 \\
w/ Language Only
 & .0621 & .0957 & .0335 & .0421
 & .0712 & .1081 & .0389 & .0484
 & .0618 & .0925 & .0337 & .0414
 & .0293 & .0458 & .0161 & .0206 \\
\midrule
\rowcolor[HTML]{f0f0f0}
\multicolumn{17}{c}{\textit{\textbf{Effect of Dual Alignment}}} \\
w/o $\mathcal{L}_\text{vision}$
 & .0702 & .1085 & .0379 & .0477
 & .0801 & .1221 & .0438 & .0546
 & .0701 & .1049 & .0383 & .0471
 & .0333 & .0519 & .0183 & .0234 \\
w/o $\mathcal{L}_\text{text}$
 & .0684 & .1051 & .0369 & .0464
 & .0778 & .1186 & .0425 & .0530
 & .0682 & .1019 & .0372 & .0457
 & .0324 & .0503 & .0178 & .0227 \\
w/o Both
 & .0658 & .1012 & .0354 & .0445
 & .0752 & .1148 & .0411 & .0513
 & .0659 & .0985 & .0359 & .0441
 & .0311 & .0486 & .0170 & .0218 \\
\midrule
\rowcolor[HTML]{f0f0f0}
\multicolumn{17}{c}{\textit{\textbf{Choice of External Frozen Vision Encoder}}} \\
w/ DINOv2
 & .0715 & .1103 & .0386 & .0485
 & .0811 & .1238 & .0444 & .0554
 & .0710 & .1064 & .0387 & .0477
 & .0338 & .0528 & .0186 & .0238 \\
\midrule
\rowcolor[HTML]{f0f0f0}
\multicolumn{17}{c}{\textit{\textbf{Effect of Latent Reasoning}}} \\
w/o Latent Reasoning
 & .0612 & .0948 & .0330 & .0416
 & .0695 & .1058 & .0380 & .0474
 & .0598 & .0895 & .0326 & .0400
 & .0278 & .0435 & .0152 & .0195 \\
\midrule
\rowcolor[HTML]{ecf0ff}
\textbf{\textsc{LARK}}
 & \textbf{.0741} & \textbf{.1148} & \textbf{.0402} & \textbf{.0506}
 & \textbf{.0842} & \textbf{.1285} & \textbf{.0462} & \textbf{.0576}
 & \textbf{.0738} & \textbf{.1108} & \textbf{.0403} & \textbf{.0498}
 & \textbf{.0352} & \textbf{.0548} & \textbf{.0194} & \textbf{.0248} \\
\bottomrule
\end{tabular}
}
\end{table*}

\subsection{Ablation Study and Analysis}
\minisection{Effect of CoT Supervision.}
Removing CoT supervision (w/o CoT) leads to a consistent degradation across all datasets (e.g., R@20: $-$6.1\% on Baby), confirming that intermediate language supervision helps the latent tokens converge to semantically meaningful hidden states, even though the language head can be bypassed at inference.
The Language Only variant, which replaces latent tokens with discrete text generation, incurs a further drop ($-$9.5\% on Baby).
The gap between w/o CoT and Language Only isolates the benefit of continuous representations: even without CoT guidance, latent tokens outperform discrete tokens because they bypass the vocabulary bottleneck and retain fine-grained information that quantization to a finite token set would destroy.

\minisection{Effect of Dual Alignment.}
We ablate the two alignment losses independently.
Removing the vision alignment $\mathcal{L}_\text{vision}$ causes a moderate drop (e.g., R@20: $-$5.5\% on Baby), indicating that anchoring latent representations to the frozen vision encoder during Stage~1 helps preserve visual fidelity across reasoning steps.
Removing the reasoning-text alignment $\mathcal{L}_\text{text}$ yields a larger degradation (e.g., R@20: $-$8.4\% on Baby), confirming that grounding Stage~2 features to the CoT hidden states is essential for retaining the reasoning semantics in the final item embeddings.
When both losses are removed, performance degrades further (e.g., R@20: $-$11.8\% on Baby), demonstrating that the two alignment objectives are complementary: $\mathcal{L}_\text{vision}$ preserves visual information and $\mathcal{L}_\text{text}$ preserves reasoning conclusions, jointly counteracting cross-modal dilution across the two stages.

\minisection{Effect of Latent Reasoning.}
To disentangle the contribution of latent reasoning from the stronger VLM backbone, we remove all latent tokens and CoT supervision, directly pooling the VLM's last hidden states for contrastive training (w/o Latent Reasoning).
Results in Table~\ref{tab:ablation} show that the VLM backbone alone already outperforms most baselines from Table~\ref{tab:main}, but still underperforms \textsc{LARK} by $-$17.4\% R@20 on Baby, confirming that the interleaved latent reasoning mechanism is a substantial source of gains beyond the backbone itself.

\subsection{Hyperparameter Sensitivity}
\label{sec:hyperparam}
\begin{figure}[h]
\centering
\includegraphics[width=\linewidth]{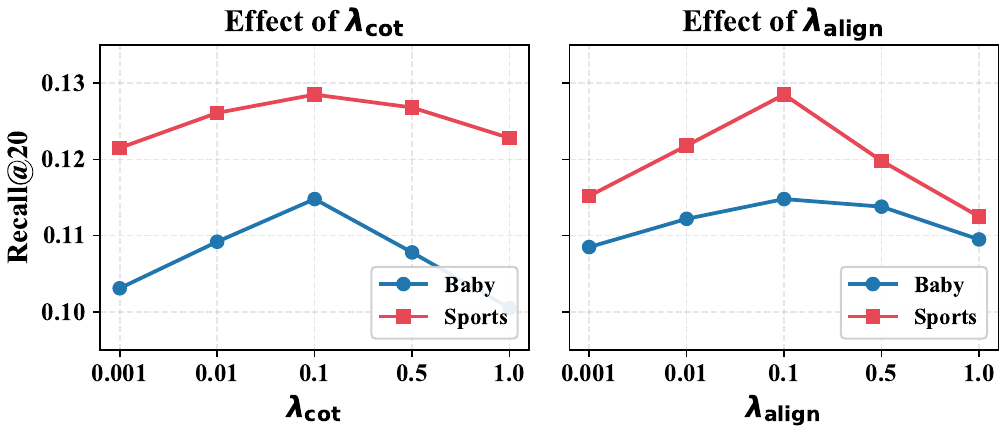}
\vspace{-6mm}
\caption{Effect of loss weights $\lambda_\text{cot}$ and $\lambda_\text{align}$ on Recall@20.}
\label{fig:hyperparam}
\end{figure}

\minisection{Effect of Loss Weights.}
Figure~\ref{fig:hyperparam} shows the effect of $\lambda_\text{cot}$ and $\lambda_\text{align}$ on Recall@20 across Baby and Sports.
Both exhibit an inverted-U shape peaking around 0.1, indicating that \textsc{LARK} is robust within a reasonable range.
Excessively large weights ($\geq 0.5$) cause the auxiliary objectives to dominate the contrastive loss, degrading embedding quality.

\begin{figure}[h]
\centering
\includegraphics[width=\linewidth]{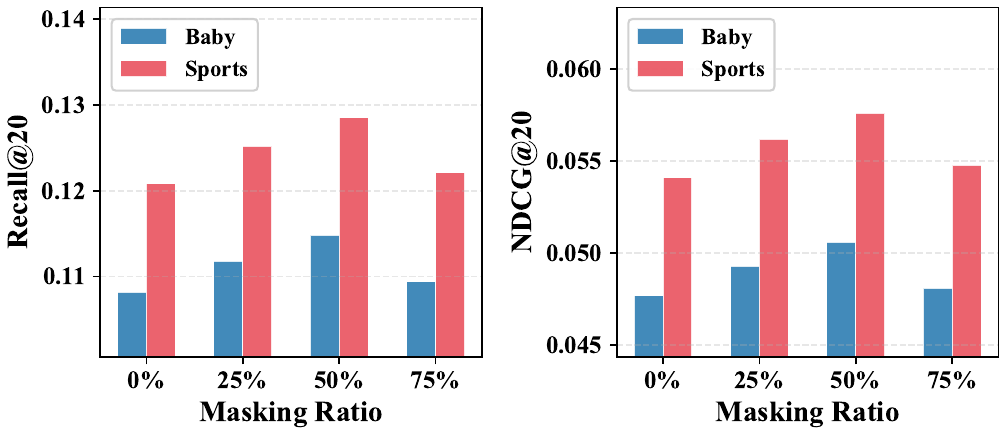}
\vspace{-7mm}
\caption{Effect of masking ratio $\rho$ on Recall@20 and NDCG@20.}
\label{fig:masking}
\end{figure}

\minisection{Effect of Masked Vision Encoder.}
Figure~\ref{fig:masking} reports performance under different masking ratios $\rho \in \{0\%, 25\%, 50\%, 75\%\}$ on Baby and Sports.
A ratio of 50\% yields the best results: without masking, the model tends to overfit to pixel-level details, while excessive masking discards too much visual information for coherent representations.

\begin{figure}[h]
\centering
\includegraphics[width=0.73\linewidth]{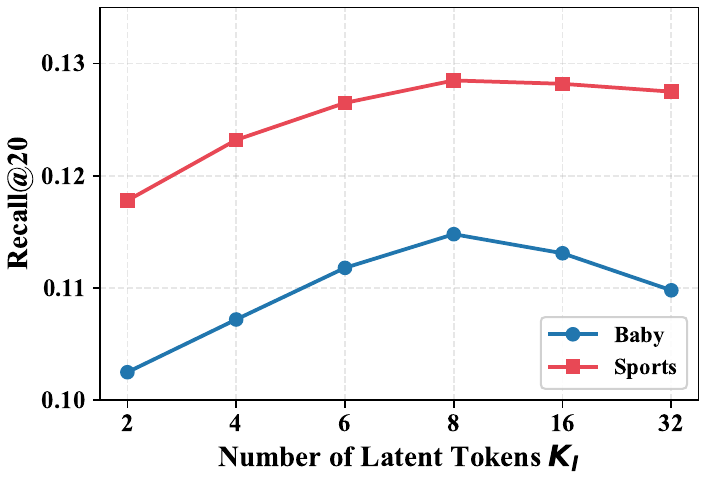}
\vspace{-4mm}
\caption{Effect of total latent token count $K_l = N \times K$ on Recall@20.}
\label{fig:num_latent}
\end{figure}

\minisection{Effect of Reasoning Length.}
Figure~\ref{fig:num_latent} examines the effect of $K_l \in \{4, 8, 12, 16, 24, 32\}$ on Baby and Sports, where $K_l = N \times K$ is the total number of latent tokens.
Performance improves from $K_l = 4$ to $K_l = 16$ and plateaus beyond $K_l = 24$.
Increasing to $K_l = 32$ yields marginal gains while doubling the computational cost of the first stage, so we set $K_l = 16$ ($N=4, K=4$) as the default.

\minisection{Choice of External Frozen Vision Encoder.}
We compare different choices for the frozen encoder used in the vision alignment loss.
Using the VLM's own ViT as the alignment target yields the best performance across all datasets, as it shares the same representation space as the trainable pathway, allowing the alignment signal to directly regularize the latent features without cross-architecture mismatch.
Replacing it with DINOv2~\citep{oquab2024dinov2}, a strong self-supervised vision encoder, still improves over removing vision alignment entirely (cf.\ Table~\ref{tab:ablation}, w/o $\mathcal{L}_\text{vision}$), but underperforms the default setting by a consistent margin (e.g., R@20: $-$3.9\% on Baby).
This gap suggests that the distributional shift between DINOv2's feature space and the VLM's internal representations introduces a misalignment that partially offsets the regularization benefit.
Notably, this design also simplifies deployment, as no additional external encoder needs to be loaded -- the frozen copy is derived from the VLM itself.

\subsection{Cross-Modal Attention Analysis}
\label{sec:dilution}
To empirically verify the cross-modal dilution phenomenon motivating our dual alignment design, we visualize how attention to image and text tokens evolves across reasoning steps.
For each generation step in Stage~1, we compute the average attention weight that the current token assigns to image tokens and text tokens, respectively, at the middle transformer layer.
Figure~\ref{fig:attention_decay} compares \textsc{LARK} (with dual alignment) against a variant without alignment losses.
Without alignment, attention to both image and text tokens decays rapidly as reasoning progresses: image attention drops by $\sim$45\% and text attention by $\sim$38\% from the first to the last reasoning step, confirming that cross-modal signals attenuate during multi-step reasoning~\citep{jeon2026vision}.
With \textsc{LARK}'s dual alignment, attention to both modalities remains substantially more stable (image attention drops by only $\sim$12\%, text attention by $\sim$9\%), demonstrating that the vision alignment loss $\mathcal{L}_\text{vision}$ and reasoning-text alignment loss $\mathcal{L}_\text{text}$ effectively counteract this dilution.
We further verify this at the representation level through CKA similarity analysis in \S\ref{sec:drift}, confirming that cross-modal dilution manifests as genuine embedding drift, not merely an attention-level artifact.

\subsection{In-depth Analysis}
\label{sec:indepth}
\minisection{Item-to-Item Recommendations.}
Since NoteLLM-2~\citep{zhang2025notellm} is originally designed for item-to-item (i2i) recommendation, we conduct a direct comparison on its native task using the Baby and Sports datasets.
For each item, we construct ground-truth similar items from co-interaction patterns (item pairs sharing $\geq 5$ common users) using held-out test-set interactions, which are disjoint from the training-set interactions used to build the Swing graph, ensuring no overlap between supervision and evaluation.
Table~\ref{tab:i2i} shows that \textsc{LARK} outperforms NoteLLM-2 by 4.7\% and 4.3\% on R@20 for Baby and Sports, respectively, even on the i2i recommendation task where NoteLLM-2 was originally designed.
Notably, the performance gap is narrower compared to user-to-item recommendation (cf.\ Table~\ref{tab:main}), confirming NoteLLM-2's i2i-native design advantage.
Nevertheless, \textsc{LARK}'s two-stage latent reasoning captures richer item-level semantics that transfer well to direct item similarity computation, yielding consistent improvements across both datasets.

\begin{table}[t]
\centering
\caption{Performance comparison of item-to-item recommendation task on Baby and Sports.}
\label{tab:i2i}
\vspace{-3mm}
\resizebox{0.65\columnwidth}{!}{
\small
\begin{tabular}{l cccc}
\toprule
\multirow{2}[2]{*}{\textbf{Method}}
  & \multicolumn{2}{c}{\textbf{Baby}}
  & \multicolumn{2}{c}{\textbf{Sports}} \\
\cmidrule(lr){2-3}\cmidrule(lr){4-5}
 & R@10 & R@20 & R@10 & R@20 \\
\midrule
LightGCN & .0542 & .0828 & .0638 & .0972 \\
VBPR & .0598 & .0912 & .0705 & .1072 \\
FREEDOM & .0672 & .1025 & .0791 & .1205 \\
AlignRec & .0715 & .1088 & .0842 & .1278 \\
NoteLLM-2 & \underline{.0738} & \underline{.1122} & \underline{.0868} & \underline{.1318} \\
\textsc{LARK} & \textbf{.0772} & \textbf{.1175} & \textbf{.0905} & \textbf{.1375} \\
\bottomrule
\end{tabular}
}
\end{table}

\begin{table}[t]
\centering
\caption{Replacing original features with \textsc{LARK} embeddings.}
\label{tab:feature_replace}
\resizebox{\columnwidth}{!}{
\small
\begin{tabular}{l cccc cccc}
\toprule
\multirow{2}[2]{*}{\textbf{Method}}
  & \multicolumn{4}{c}{\textbf{Baby}}
  & \multicolumn{4}{c}{\textbf{Sports}} \\
\cmidrule(lr){2-5}\cmidrule(lr){6-9}
 & \multicolumn{2}{c}{R@20} & \multicolumn{2}{c}{N@20}
 & \multicolumn{2}{c}{R@20} & \multicolumn{2}{c}{N@20} \\
\cmidrule(lr){2-3}\cmidrule(lr){4-5}\cmidrule(lr){6-7}\cmidrule(lr){8-9}
 & Orig. & +\textsc{LARK} & Orig. & +\textsc{LARK}
 & Orig. & +\textsc{LARK} & Orig. & +\textsc{LARK} \\
\midrule
MMGCN   & .0660 & .0752 & .0282 & .0321 & .0636 & .0732 & .0270 & .0310 \\
LATTICE & .0850 & .0935 & .0370 & .0405 & .0953 & .1045 & .0421 & .0460 \\
BM3     & .0883 & .0958 & .0383 & .0415 & .0980 & .1068 & .0438 & .0476 \\
FREEDOM & .0992 & .1048 & .0424 & .0448 & .1089 & .1152 & .0481 & .0510 \\
\bottomrule
\end{tabular}
}
\end{table}

\minisection{Feature Replacement Analysis.}
To further examine the quality of \textsc{LARK}'s learned representations, we replace the original visual and textual features used by existing multi-modal recommendation models with \textsc{LARK}'s pre-computed item embeddings, keeping all other model components unchanged.
Table~\ref{tab:feature_replace} reports R@20 and N@20 on Baby and Sports.
All four models benefit from \textsc{LARK} embeddings, with consistent improvements across both datasets.
The gains are inversely correlated with the original model's strength: MMGCN, which relies on basic CNN features, sees the largest improvement (+13.9\% R@20 on Baby, +15.1\% on Sports), while FREEDOM, which already employs sophisticated feature refinement, improves by a smaller but still meaningful margin (+5.6\% on Baby, +5.8\% on Sports).
This pattern suggests that \textsc{LARK}'s VLM-derived embeddings capture visual and semantic information that traditional feature extractors (e.g., ResNet, BERT) fail to encode, and that this information is complementary to the graph-based learning in downstream models.
Notably, even FREEDOM -- the strongest baseline in Table~\ref{tab:main} -- achieves R@20 of 0.1048 on Baby when equipped with \textsc{LARK} features, approaching \textsc{LARK}\textsubscript{LightGCN}'s 0.1148, which demonstrates that the quality of item representations is the primary driver of recommendation performance.

\begin{figure}[t]
\centering
\includegraphics[width=\linewidth]{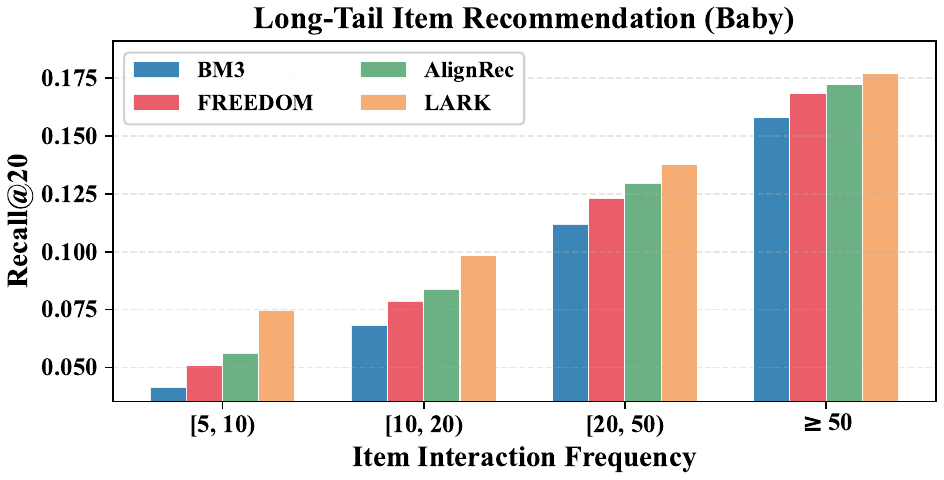}
\vspace{-5mm}
\caption{Long-tail item recommendation on Baby. \textsc{LARK} shows the largest improvement on low-frequency items.}
\label{fig:longtail}
\end{figure}

\minisection{Long-Tail Items Recommendations.}
A key challenge in multimodal recommendation is handling long-tail items that have sparse interaction histories.
Following AlignRec~\citep{liu2024alignrec}, we partition items into groups by interaction frequency and report Recall@20 per group on the Baby dataset.
As shown in Figure~\ref{fig:longtail}, \textsc{LARK} achieves substantial improvements over baselines on the lowest-frequency group (items with fewer than 5 interactions), where collaborative signals are insufficient and the model must rely on multimodal content understanding.
This validates that the multi-step latent reasoning captured by \textsc{LARK}'s dual alignment provides meaningful representations even for items with minimal behavioral data.

\minisection{Embedding Drift Analysis.}
\label{sec:drift}
Beyond the attention-level analysis in \S\ref{sec:dilution}, we directly measure how intermediate representations drift from the original multimodal features as reasoning progresses.
For each reasoning step $n \in \{1, \ldots, N\}$, we extract the hidden states at the $n$-th latent group and compute their centered kernel alignment (CKA)~\citep{kornblith2019similarity} with (i) the frozen vision encoder features $\mathbf{F}_\phi$ and (ii) the input text token embeddings $\mathbf{W}_i$.
We evaluate on 1,000 randomly sampled items from the Baby dataset and compare \textsc{LARK} (with dual alignment) against a variant without alignment losses.

\begin{figure}[t]
\centering
\includegraphics[width=\linewidth]{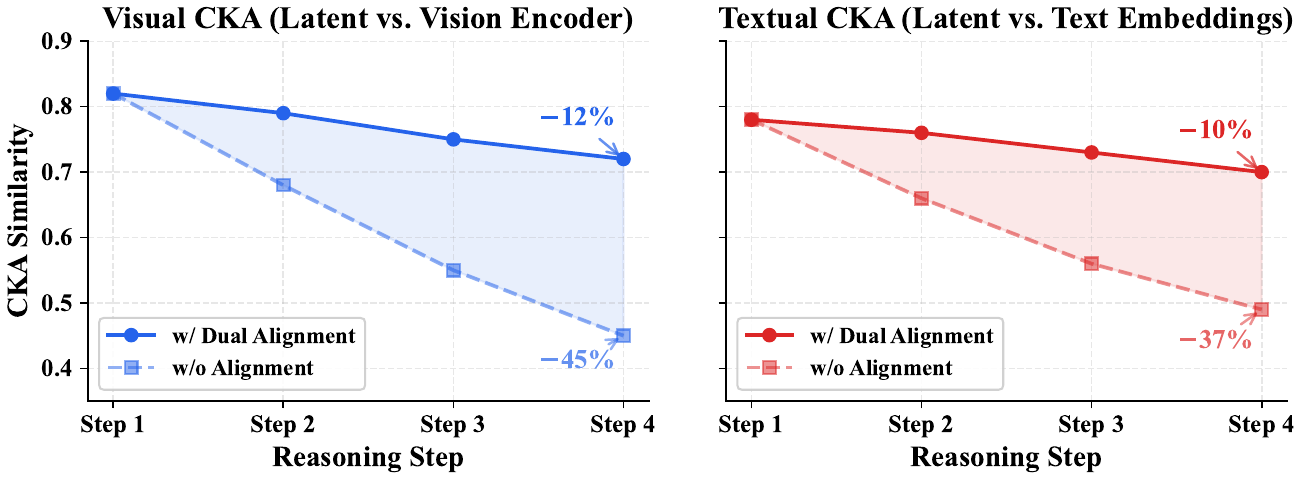}
\vspace{-5mm}
\caption{CKA similarity between latent representations and original multimodal features across reasoning steps on Baby. \textbf{Left:} visual CKA (latent vs.\ frozen vision encoder). \textbf{Right:} textual CKA (latent vs.\ input text embeddings). Without alignment, both modalities exhibit substantial representational drift; \textsc{LARK}'s dual alignment effectively preserves multimodal grounding.}
\label{fig:drift}
\end{figure}

As shown in Figure~\ref{fig:drift}, without alignment, both visual CKA and textual CKA decay monotonically across reasoning steps: visual CKA drops from 0.82 to 0.45 ($-$45\%) and textual CKA from 0.78 to 0.49 ($-$37\%) by the fourth step.
This confirms that cross-modal dilution is not merely an attention-level artifact but manifests as genuine representational drift --- the latent features progressively lose their grounding in the original multimodal inputs.
With \textsc{LARK}'s dual alignment, visual CKA remains above 0.72 throughout (dropping only $-$12\%), and textual CKA stays above 0.70 ($-$10\%), demonstrating that the alignment losses effectively anchor the representations to their multimodal origins across reasoning stages.

Notably, the textual CKA stabilization is primarily attributable to the reasoning-text alignment $\mathcal{L}_\text{text}$ in Stage~2, which back-propagates semantic anchoring signals through the Bridge MLP into the first stage's latent representations during training.

\section{Conclusion and Future Work}
We present \textsc{LARK}, a latent-aligned reasoning framework that repurposes a vision-language model as a multimodal item encoder for recommendation.
\textsc{LARK} interleaves latent tokens with multi-step chain-of-thought reasoning in a fixed template, and applies dual alignment -- vision alignment in Stage~1 and reasoning-text alignment in Stage~2 -- to counteract cross-modal dilution across the reasoning pipeline.
Experiments on three Amazon benchmarks and one industrial dataset demonstrate consistent improvements over state-of-the-art baselines, with up to 15.1\% gains as drop-in feature replacements.
As future work, we plan to deploy \textsc{LARK} in a production environment for online A/B evaluation and to extend the framework to user-side modeling by incorporating interaction history into the latent reasoning process.

\clearpage
\bibliographystyle{ACM-Reference-Format}
\bibliography{main}
\clearpage
\appendix

\section{Overall Algorithm}
\label{app:algo}

We summarize the complete training and inference procedures of \textsc{LARK} in Algorithm~\ref{alg:lark_train} and Algorithm~\ref{alg:lark_infer}, respectively.
Training alternates between Stage~1 (interleaved latent-language reasoning with vision alignment) and Stage~2 (item-item contrastive learning with reasoning-text alignment).
At inference, the same two-stage pipeline produces the final item embedding $\mathbf{e}_i$ for each item.

\begin{algorithm}[h]
	\caption{Training of \textsc{Lark}}
	\label{alg:lark_train}
	\KwIn{
		Item set $\mathcal{I}$ with images $\{\mathbf{I}_i\}$ and texts $\{\mathbf{T}_i\}$;
		Swing positive sets $\{\mathcal{N}_i^+\}$;
		Multi-step CoT annotations $\{\mathbf{c}_i^*\}$ ($N$ steps);
		VLM with vision encoder $\mathcal{F}_{\text{vision}}$, frozen copy $\mathcal{F}_{\text{frozen}}$, transformer $\mathcal{F}_{\text{transformer}}$, Bridge MLP $\mathcal{F}_\text{BridgeMLP}$, projection head $\mathcal{F}_\text{ProjectionHead}$
	}
	\vspace{1mm}
	Initialize all parameters\;
	\Repeat{convergence}{
		Sample a batch $\mathcal{B}$ from $\mathcal{I}$\;
		\ForEach{item $i$ in $\mathcal{B}$}{
			\tcp{Stage 1: Vision-Language Reasoning}
			$\mathbf{v}_i \leftarrow \mathcal{F}_{\text{vision}}(\text{Mask}(\mathbf{I}_i, \rho))$\;
			$\mathbf{S} \leftarrow [\mathbf{V}_i;\, \mathbf{W}_i;\, \mathbf{P}^{(1)}]$\;
			\For(\tcp*[f]{$N$ groups}){$n = 1$ \KwTo $N$}{
				\For(\tcp*[f]{$K$ latent tokens}){$k = 1$ \KwTo $K$}{
					$h \leftarrow \mathcal{F}_{\text{transformer}}(\mathbf{S})[-1,:]$;\quad $\mathbf{S} \leftarrow [\mathbf{S};\, h]$\;
				}
				\For(\tcp*[f]{teacher-forced CoT}){each $\bm{c}_t^*$ in step $n$}{
					$\mathbf{S} \leftarrow [\mathbf{S};\, \mathcal{F}_\text{embed}(\bm{c}_t^*)]$\;
				}
			}
			Collect $\mathbf{H}^{(1)}_i$ from $K_l{=}N{\times}K$ latent positions, $\mathbf{H}^{(1)}_{\text{lang},i}$ from language positions\;
			$\hat{\mathbf{F}} \leftarrow \mathcal{F}_\text{up}(\mathbf{H}^{(1)}_i)$;\; $\mathbf{F}_\phi \leftarrow \mathcal{F}_{\text{frozen}}(\mathbf{I}_i)$ \tcp*{for $\mathcal{L}_{\text{vision}}$}
			\tcp{Stage 2: Item-Item Alignment}
			$\tilde{\mathbf{H}}^{(1)}_i \leftarrow \mathcal{F}_\text{BridgeMLP}(\mathbf{H}^{(1)}_i)$;\quad $\mathbf{E}^{(2)} \leftarrow [\tilde{\mathbf{H}}^{(1)}_i;\, \mathbf{p}^{(2)}]$\;
			$\mathbf{H}^{(2)}_i \leftarrow \mathcal{F}_{\text{transformer}}(\mathbf{E}^{(2)})$;\quad $\mathbf{e}_i \leftarrow \mathcal{F}_\text{ProjectionHead}(\mathbf{H}^{(2)}_i)$\;
			Extract $\mathbf{F}^{(2)}_i$ at layer $m$;\; $\bar{\mathbf{r}} \leftarrow \text{MeanPool}(\mathbf{H}^{(1)}_{\text{lang},i})$ \tcp*{for $\mathcal{L}_{\text{text}}$}
		}
		Update by $\mathcal{L} = \mathcal{L}_{\text{i2i}} + \lambda_\text{cot} \mathcal{L}_{\text{cot}} + \lambda_\text{align} (\mathcal{L}_{\text{vision}} + \mathcal{L}_{\text{text}})$\;
	}
\end{algorithm}

\begin{algorithm}[h]
	\caption{Inference of \textsc{Lark}}
	\label{alg:lark_infer}
	\KwIn{Item set $\mathcal{I}$; trained VLM, Bridge MLP $\mathcal{F}_\text{BridgeMLP}$, projection head $\mathcal{F}_\text{ProjectionHead}$}
	\KwOut{Item embeddings $\{\mathbf{e}_i\}_{i \in \mathcal{I}}$}
	\vspace{1mm}
	\ForEach{item $i$ in $\mathcal{I}$}{
		$\mathbf{S} \leftarrow [\mathbf{V}_i;\, \mathbf{W}_i;\, \mathbf{P}^{(1)}]$\;
		\For{$n = 1$ \KwTo $N$}{
			\For{$k = 1$ \KwTo $K$}{
				$h \leftarrow \mathcal{F}_{\text{transformer}}(\mathbf{S})[-1,:]$;\quad $\mathbf{S} \leftarrow [\mathbf{S};\, h]$\;
			}
			Autoregressively generate CoT tokens for step $n$; append to $\mathbf{S}$\;
		}
		Collect $\mathbf{H}^{(1)}_i$ from latent positions\;
		$\mathbf{e}_i \leftarrow \mathcal{F}_\text{ProjectionHead}(\mathcal{F}_{\text{transformer}}([\mathcal{F}_\text{BridgeMLP}(\mathbf{H}^{(1)}_i);\, \mathbf{P}^{(2)}]))$\;
	}
	Store $\{\mathbf{e}_i\}$ in vector index; train downstream recommendation model
\end{algorithm}

\section{Prompts}
\label{app:prompts}
As introduced in \S\ref{sec:training}, we use Gemini-2.5-Flash~\citep{gemini2025flash} as the teacher VLM to generate multi-step chain-of-thought (CoT) reasoning annotations for all items offline.
For each item $i$, we provide its product image $\mathbf{I}_i$ and title text $\mathbf{T}_i$ as input, and prompt the model to produce a structured four-step reasoning chain covering visual attributes, functional analysis, target audience, and distinguishing features.
The generation is performed in a single pass over the entire item catalog before training begins.

The prompt template used for CoT generation is shown below:

\begin{tcolorbox}[colback=gray!5, colframe=gray!50, title=CoT Generation Prompt, fonttitle=\bfseries\small, boxrule=0.5pt, arc=2pt]
\small
\texttt{You are a product understanding assistant. Given the product image and title, generate a structured analysis following these four steps:}

\vspace{1mm}
\texttt{Step 1 - Visual Attributes: Describe the key visual properties (material, color, shape, texture).}

\texttt{Step 2 - Functional Analysis: Identify the product's primary function and usage scenario.}

\texttt{Step 3 - Target Audience: Determine the intended user group based on visual and functional cues.}

\texttt{Step 4 - Distinguishing Features: Highlight what makes this product unique compared to similar items.}

\vspace{1mm}
\texttt{Keep each step to one sentence (under 15 words). Focus on the product itself.}

\vspace{1mm}
\texttt{Title: \{item\_title\}}

\texttt{Analysis:}
\end{tcolorbox}

\begin{table}[h]
\centering
\caption{Examples of multi-step CoT reasoning generated by Gemini-2.5-Flash.}
\label{tab:cot_examples}
\vspace{-3mm}
\resizebox{\columnwidth}{!}{
\begin{tabular}{p{1.8cm}p{2.2cm}p{6cm}}
\toprule
\textbf{Category} & \textbf{Title (Excerpt)} & \textbf{Generated CoT Reasoning} \\
\midrule
Baby & Infant Soft Cotton Romper &
\textit{Visual:} Pastel cotton romper with snap buttons. \textit{Function:} Easy-change infant daily wear. \textit{Audience:} Parents of newborns 0--12 months. \textit{Distinguishing:} Snap-button design for quick diaper access. \\
\addlinespace
Sports & Adjustable Dumbbell Set 25lb &
\textit{Visual:} Black rubber-coated dumbbell with chrome handle. \textit{Function:} Adjustable weight for home strength training. \textit{Audience:} Home fitness enthusiasts at beginner-intermediate level. \textit{Distinguishing:} Quick-lock weight adjustment mechanism. \\
\addlinespace
Clothing & Women's Knit Cardigan &
\textit{Visual:} Cream lightweight knit with ribbed cuffs. \textit{Function:} Layering piece for casual or office settings. \textit{Audience:} Women seeking versatile everyday outerwear. \textit{Distinguishing:} Open-front design with elongated silhouette. \\
\bottomrule
\end{tabular}
}
\end{table}

Table~\ref{tab:cot_examples} provides representative examples of generated CoT reasoning chains across different product categories.

\begin{table}[h]
\centering
\caption{Statistics of generated CoT reasoning chains.}
\label{tab:cot_stats}
\vspace{-3mm}
\begin{tabular}{lrrrr}
\toprule
\textbf{Dataset} & \textbf{\#Items} & \textbf{Avg.\ Len.} & \textbf{Med.\ Len.} & \textbf{Max Len.} \\
\midrule
Baby     & 7,050   & 53.2 & 54 & 72 \\
Sports   & 18,357  & 56.8 & 56 & 75 \\
Clothing & 23,033  & 54.1 & 53 & 70 \\
In-House & 432,671 & 55.4 & 55 & 74 \\
\bottomrule
\end{tabular}
\end{table}

Table~\ref{tab:cot_stats} summarizes the statistics of the generated CoT data.
The average reasoning chain length is approximately 55 tokens (across four steps), consistent with our design goal of structured multi-step reasoning that provides rich supervision for latent token training.

We apply the following post-processing steps to ensure data quality:
\begin{enumerate}[leftmargin=*,nosep]
    \item \textbf{Step verification}: Reasoning chains missing any of the four steps are re-generated; fewer than 1.2\% of items require re-generation.
    \item \textbf{Length filtering}: Individual steps exceeding 15 words are truncated at the last complete word.
    \item \textbf{Deduplication}: Identical reasoning chains across different items are flagged; fewer than 0.3\% of items produce duplicates.
    \item \textbf{Format normalization}: All text is lowercased and stripped of trailing punctuation for consistency.
\end{enumerate}

\end{document}